\documentclass[final,10pt,twocolumn,5p]{article}

\usepackage{graphicx}

\usepackage{amssymb}

\usepackage{amsmath}
\usepackage{hyperref}
\newcommand{\mean}[1]{\ensuremath{\la #1 \ra}}
\newcommand{\deriv}[2]{\ensuremath{\frac{\partial#1}{\partial#2}}}

\newcommand{\eq}{\ensuremath{\epsilon_q}}

\newcommand{\la}{\ensuremath{\langle}}
\newcommand{\ra}{\ensuremath{\rangle}}
\newcommand{\fn}{\ensuremath{\vec{f}_{n}}}
\newcommand{\ft}{\ensuremath{\vec{f}_{t}}}
\newcommand{\ri}{\ensuremath{\vec{r}_{i}}}
\newcommand{\rj}{\ensuremath{\vec{r}_{j}}}
\newcommand{\nij}{\ensuremath{\hat{n}_{ij}\,}}
\newcommand{\hij}{\ensuremath{h_{ij}}}
\newcommand{\vij}{\ensuremath{\vec{v}_{ij}}}
\newcommand{\vi}{\ensuremath{\vec{v}_{i}}}
\newcommand{\vj}{\ensuremath{\vec{v}_{j}}}

\newcommand{\mus}{\ensuremath{\mu_s}}
\newcommand{\mud}{\ensuremath{\mu_d}}
\newcommand{\vxi}{\ensuremath{\vec{\xi}\,}}
\newcommand{\ftemp}{\ensuremath{\vec{f}_t^0\,}}
\newcommand{\mij}{\ensuremath{m_{ij}}}
\newcommand{\vt}{\ensuremath{\vec{v}_{t}}}
\newcommand{\dt}{\ensuremath{\delta t}}
\newcommand{\tij}{\ensuremath{\hat{t}_{ij}\,}}

\begin{document}




\title{Influence of Rotations on the Critical State
  of Soil Mechanics}




\author{
  W.~F.~Oquendo$^{1,3}$, J.~D.~Mu\~noz$^{1,3}$ and A.~Lizcano$^{2,3}$ \\
  $^1$\small  Simulation of Physical Systems Group, Department of Physics\\
  \small  Universidad Nacional de
  Colombia, Carrera 30 No. 45-03, Ed. 404, Of. 348, Bogota DC,
  Colombia\\
  $^2$\small Department of Civil and Environmental
  Engineering\\
  \small  Universidad de los Andes, Carrera 1 No. 18A-10, Bogota
  DC, Colombia\\
  $^3$\small  CEIBA Complex Systems Research Center\\
  \small  CEiBA-Complejidad, Carrera 1 No. 18A-10, Bogota DC, Colombia
}
\maketitle



\begin{abstract}
  The ability of grains to rotate can play a crucial role on the
  collective behavior of granular media. It has been observed in
  computer simulations that imposing a torque at the contacts modifies
  the force chains, making support chains less important. In this work
  we investigate the effect of a gradual hindering of the grains
  rotations on the so-called critical state of soil mechanics. The
  critical state is an asymptotic state independent of the initial
  solid fraction where deformations occur at a constant shear strength
  and compactness. We quantify the difficulty to rotate by a friction
  coefficient at the level of particles, acting like a threshold. We
  explore the effect of this particle-level friction coefficient on
  the critical state by means of molecular dynamics simulations of a
  simple shear test on a poly-disperse sphere packing. We
  found that the larger the difficulty to rotate, the larger the final
  shear strength of the sample. Other micro-mechanical variables, like
  the structural anisotropy and the distribution of forces, are also influenced
  by the threshold. These results reveal the key role of rotations on
  the critical behavior of soils and suggest the inclusion of
  rotational variables into their constitutive equations.
\end{abstract}





\section{Introduction}\label{sec:intro}
Granular media is the second most used material by mankind, only after
water~\cite{degennes, gennespowderbook}. The annual production reaches
almost 10 billions metric-ton/year. The industrial processing of these
raw materials consumes nearly 10\% of the total electric energy on
Earth. Therefore, any optimization on the processing and any improve
on the knowledge of GM will have a direct impact both on economy and
welfare. Despite its ubiquitous nature, granular materials are far
from being fully understood, mainly because of the numerous and
different behaviors that can be generated collectively from the
interactions of the particular grains. Granular media can behave as a
liquid, solid or gas, and its particular behavior depends on the way
the system is stimulated.
 
Soils represent a particular and important example of granular
materials.  They are the main subject of study for civil engineers
dealing with settlements of buildings, earth pressure against
retaining walls, stability of slopes and embankments, etc. Despite the
rich complexity found in soils, like ratcheting and creep, there is a
particular state that is independent of soil consolidation history,
initial density and sample preparation: the critical
state~\cite{casagrandeCritical, AAPenaCritical, frictionRotations,
  hypo05, hinrichsen2004physicsGranMedia, contactMechCritical}. In
this steady state the deformation occurs at a constant average volume,
void ratio, and effective stresses. The critical state has been
characterized and studied from long time; however, its microscopic
origins remain elusive.

The critical state is a fundamental concept in soil mechanics.  It is
linked with the so-called critical state constitutive models and
failure criteria.  Since the seminal work of Casagrande in
1936~\cite{casagrandeCritical}, it has been found in sands and other
types of granular media~\cite{casagrandeCritical, AAPenaCritical,
  frictionRotations, hypo05, shapeH, bookPhysGranMedia,
  bookGranular02}. It is possible to reach the critical state through
shearing, bi-axial or tri-axial tests.  It is independent of the initial
density, sample preparation or even shear rate, provided one is
working in the quasi-static approximation.

The values of the macroscopic variables characterizing the critical
state depend both on external and internal variables. External
control variables are the confining pressure and the shear velocity;
internal variables are the grains' size distribution and shape, the
grain-to-grain friction coefficients, etc. It has been observed that
the shape of the particles and its rotational freedom affect the final
steady values~\cite{AAPenaCritical, frictionRotations,
  azema2007ForceTrans}.  In contrast with more complex shapes, spheres
(or discs) favor rotations, because there is no geometric
interlocking among them. If rotations are suppressed for spheres in
some way, it is expected that the effective shear strength will
increase.  Suiker and Flerk ~\cite{frictionRotations} showed by
three-dimensional computer simulations with spheres that a complete
hindering of rotations modifies the critical shear strength.
Similarly, A.~A.~Pe\~na, R. Garc\'{\i}a-Rojo and
H.~J.~Herrmann~\cite{AAPenaCritical, shapeH} studied the effect of
particle shape on the critical state by using two particle kinds in
2D: almost isotropic and elongated. They found that both the void
ratio and the coordination number at the critical state are different
(among other variables) for isotropic and elongated particles.
Nevertheless, no systematic study has been done regarding the role of
rotations as a continuous parameter, for instance, by hindering
gradually the rotations.

The main goal of the present work is to further explore the effect of
restricting the rotation of the spheres on the macroscopic variables
characterizing the critical state. This restriction is set by imposing
a rotation threshold for each grain. The threshold is just the sum of
the modulus of the normal forces on the grain times a rotational
coefficient, which is the same for all grains.  This procedure allows
us to mimic the effect of geometric interlocking, but increasing the
hindering to rotate in a continuous way. In other words, we are
simulating the effect of complex shapes by adding a rotation
restriction on rounded grains. The effective shear strength, the mean
coordination number, the effective stresses, and the force and torque
distributions, will be investigated as a function of this rotational
threshold. The study will be performed through molecular dynamics
simulations of a bi-dimensional and periodic poly-disperse system
under simple shear.
Section~\ref{sec:criticalState} introduces the critical state and the
computational setup. Section~\ref{sec:dem} describes the
discrete-element model, that is the interactions among particles and
the integration method for the equations of
movement. Section~\ref{sec:results} shows the main results and,
finally, section~\ref{sec:conclu} states the main conclusions from the
work.

\section{The Critical State}\label{sec:criticalState}
\begin{figure}[!ht]
  \centering
  \includegraphics[scale=0.32]{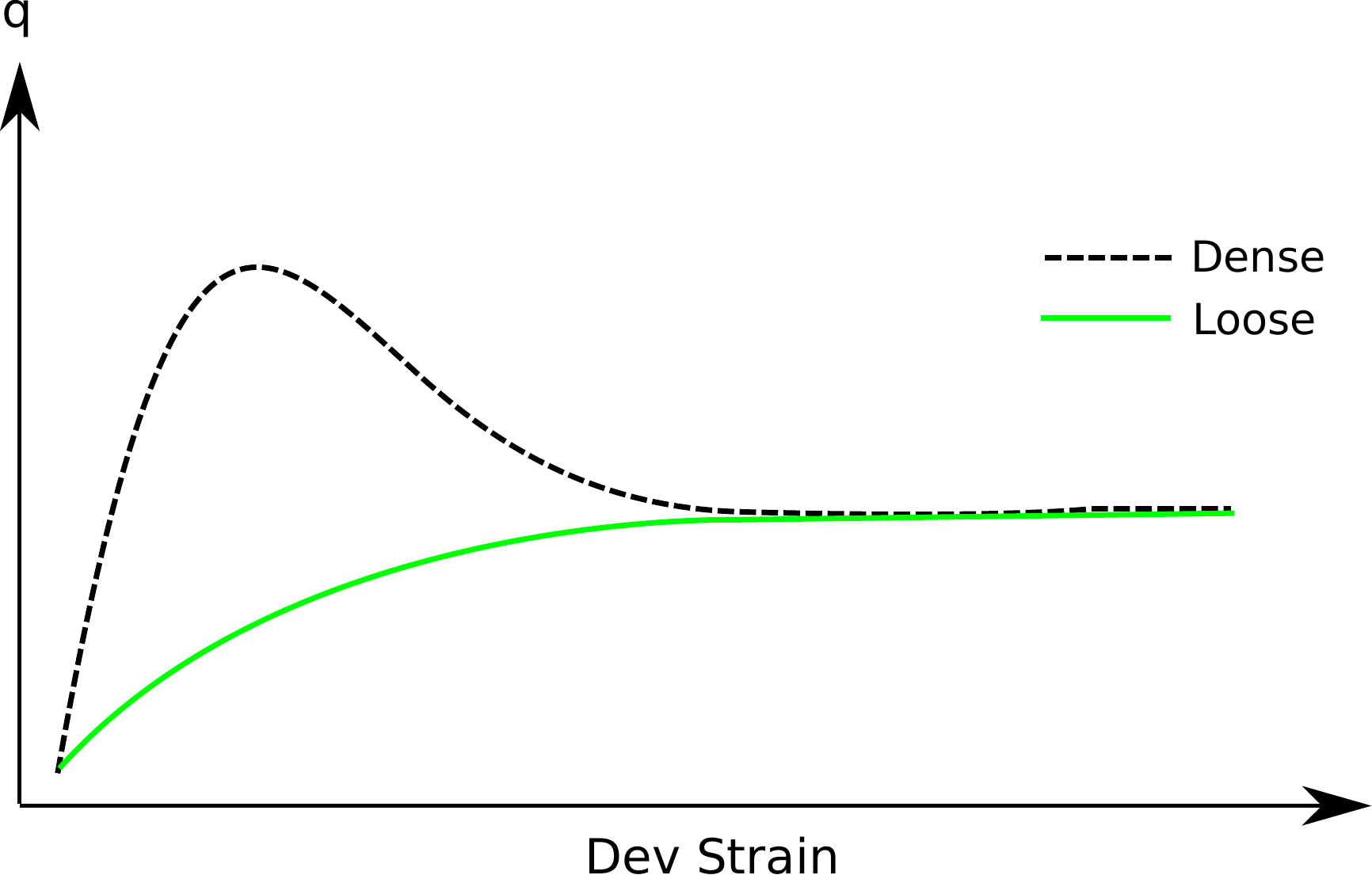} \\
  \includegraphics[scale=0.32]{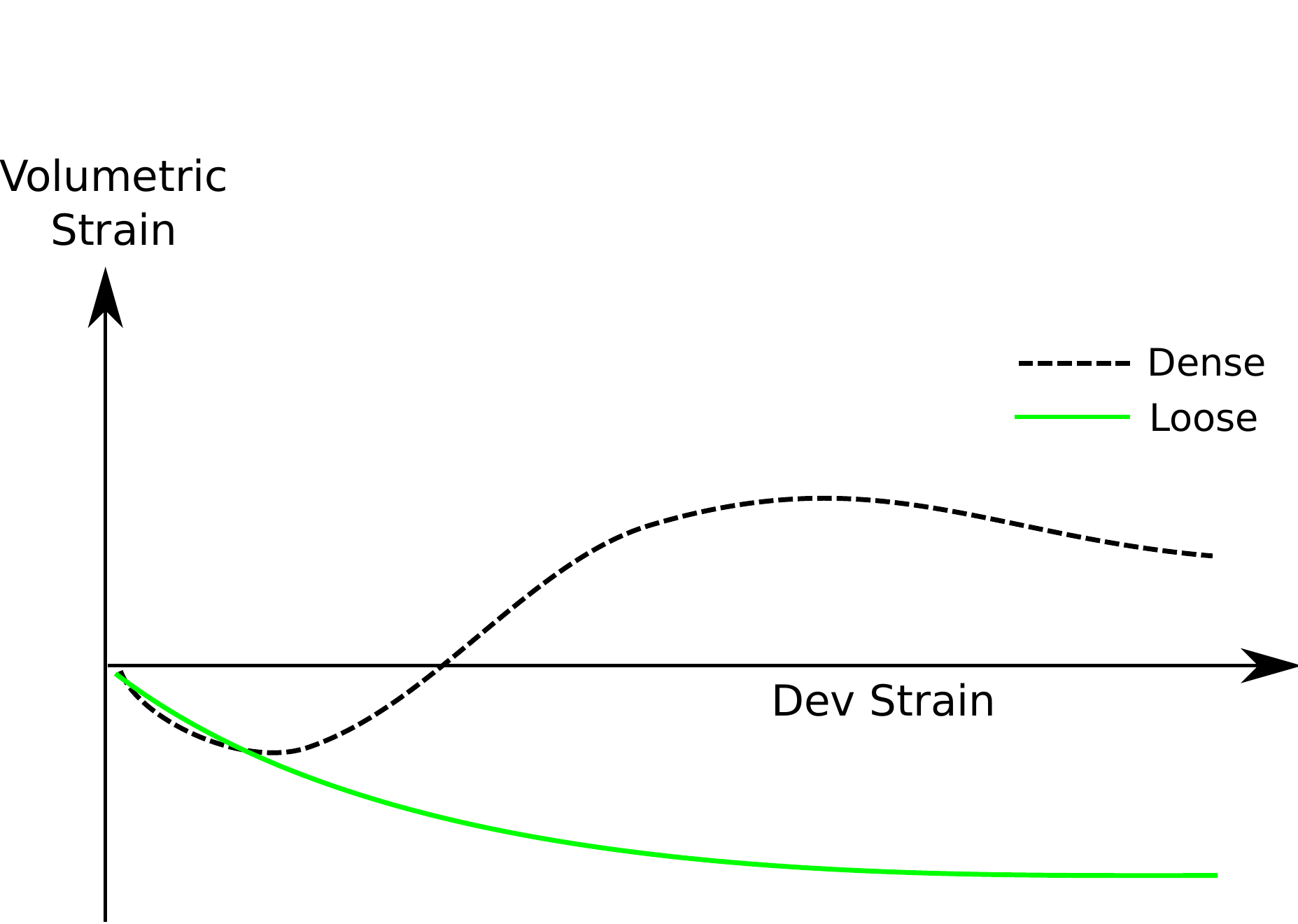}
  \caption{(Color online) (\textbf{Top}) Sketch of the typical
    responses for $q$, and (\textbf{Bottom}) of the volumetric strain,
    as functions of the deviatoric strain.} \label{fig:criticalFigs}
\end{figure}

The critical state is defined in soil mechanics as the steady
state where deformations occurs at constant average void ratio and
effective stresses~\cite{criticalBook01, criticalBook02}
(Fig.~\ref{fig:criticalFigs}).

The definition of the critical state can be expressed mathematically as
\begin{equation}\label{eq:csmathdef} 
  \deriv{\mean{e}}{t} = 
  \deriv{\mean{p}}{t} =
  \deriv{\mean{q}}{t} = 0\quad,  
\end{equation} 
where $\mean{e}$ is the average void ratio $e$$=$$V_V/V_S$ (i.e. the
ratio between voids volume $V_V$ and solids volume $V_S$), $\mean{p}$
is the average isotropic pressure, $\mean{q}$ is the average
deviatoric pressure and $\eq$ is the deviatoric strain. In 2D, the
mean pressure $p$ is computed as $p$$=$$(\sigma_1 + \sigma_3)/2$, and
the deviatoric stress as $q$$ =$$ (\sigma_1 - \sigma_3)/2 $, where
$\sigma_1$ ($\sigma_3$) is the principal value of the stress tensor on
the axial (tangential) direction.

The inertia number $I$ allows to quantify the degree of
quasi-staticity of a given test~\cite{estradaRolling, midi2004dense,
  cruz2005rheophysics, agnolin2007InternalII}, and is defined as
\begin{equation}
  \label{eq:inertiaNumber} I = \mean{d} \dot\gamma
\sqrt{\frac{\rho}{\sigma_{wall}}},
\end{equation} 
where $\dot\gamma$ is the shear rate, $\mean{d}$ is the
mean diameter, $\rho$ is the density of the solid grains and
$\sigma_{wall}$ is the confining pressure. In a quasi-static test, $I
\ll 1$.  In this work, $I \simeq 10^{-3}$. 

\section{Discrete Element Method}\label{sec:dem}
We employed the \textit{D}iscrete \textit{E}lement \textit{M}ethod
(DEM), also known as Soft Particle method or Molecular Dynamics, to
simulate the direct shear tests on two-dimensional poly-disperse
samples of spheres. Since the movements are 2D, both the translational
and rotational degrees of freedom are integrated with the Velocity
Verlet method~\cite{rapaport, poeschelBook, velverlet1, oquendoCPC}.

\begin{figure}[!ht]
 \centering
 \includegraphics[scale=0.4]{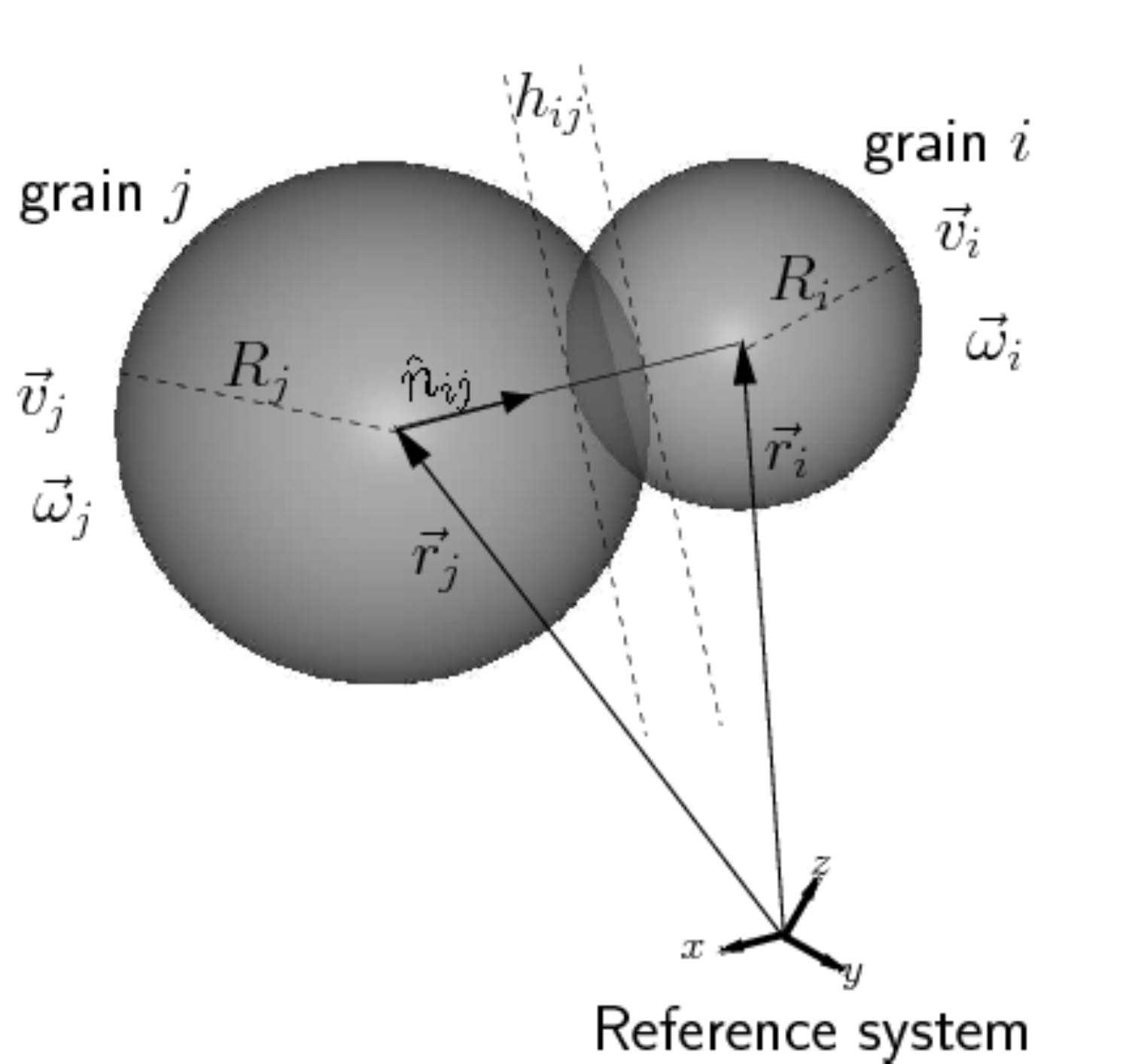}
 \caption{Relevant variables for the interaction among two spherical
   grains. } \label{fig:balls}
\end{figure}  

The normal force between two grains (see Figure \ref{fig:balls}) is
just the Hertz elastic law together with a damping term. It is given
by
\begin{equation}
 \label{eq:fn}
\fn =  \fn^{el} + \fn^{da } = k_{ij} \sqrt{R_{ij}} \hij^{3/2} \nij -\gamma m_{ij}
  (\vij\circ\nij) \nij\quad,
\end{equation} 
where $k_{ij}$ is the effective spring hardness, $R_{ij}$ is the
reduced radius, $\hij = R_i + R_j - |\ri - \rj|$ is the mutual
penetration, $\gamma$ is the damping constant in normal direction,
$m_{ij}$ is the reduced mass, $\vij$$=$$\vi - \vj$ is the relative
velocity, and
\begin{equation}
  \label{eq:nij}
  \nij = \frac{\ri - \rj}{|\ri - \rj|}\quad .
\end{equation}
 
For the tangential force we use the generalized Cundall-Strack model
proposed by Luding~\cite{ludingobjec}, with a small correction for the
tangential damping force. The algorithm is analogous for both sliding
and rolling friction, and it reads as follows: As soon as a new
contact appears (disappears) between two particles, a \textit{virtual}
spring is created (destroyed) for each of the friction processes
aforementioned. This spring stores the relative tangential
displacement. Let us denote the elongation of this tangential spring
at time $t$ by $\vxi$. The spring is always kept on the tangential
direction by means of the transformation
\begin{equation}
  \label{eq:rotspring}
  \vxi = \vxi - (\nij\circ\vxi)\nij = \nij\times\vxi\times\nij\quad.
\end{equation}
By using the current tangential spring elongation, a
temporary tangential friction force $\ftemp$ is computed as
\begin{equation}
  \label{eq:ftemp}
  \ftemp = -k_t R_{ij} \vxi - \mij \gamma_t \vt,
\end{equation}
where $k_t$ is a tangential stiffness, and $\gamma_t$ is the
tangential damping constant, which is introduced to improve the
stability of the method \cite{ludingobjec}. Static friction is present
whenever $|\ftemp|$$\le$$\mus |\fn|$; otherwise we are on the kinetic
friction regime. In the static regime $\ft$$=$$\ftemp$, and the spring
is updated as $\vxi$$=$$\vxi + \vt\dt$. In the kinetic regime, the
tangential force is given by $\ft$$=$$\mud |\fn| \tij$, where $\mud$
is the kinetic friction coefficient (typically $\mud = 0.8\mus$), and
the tangential unit vector $\tij$ is defined dynamically as
$\tij=\ftemp/|\ftemp|$. The spring is updated as
$\vxi$$=$$ \frac{-1}{k_t} (\ft + \mij \gamma_t \vt)$. The friction
torque can be computed as
\begin{equation}
  \label{eq:torquefriction}
  \tau = \vec{l}_i \times \ft,
\end{equation}
where $\vec{l}_i$ is the branch vector from the center of the particle
to the contact point. In total, four parameters are required for each
tangential interaction: $k_t, \gamma_t, \mus, \mud$; but, in practice
only $\mus$ is needed, since the other parameters are chosen
relatively to the normal force parameters and $\mus$ itself.

A main goal of the simulation is to compute the macroscopic stress
tensor $\sigma_{\alpha\beta}$. It is given
by~\cite{moreau1997macroStress, staron2005multi}
\begin{equation}
  \label{eq:stresstensor}
  \sigma_{\alpha \beta} = \frac{1}{V}\sum\limits_{c \in V} f_\alpha^c l_\beta^c\quad,
\end{equation}
where $V$ is the representative volume where the average
is done, $f_\alpha$ is $\alpha-$component of the force,
$l_\beta$ is the $\beta-$component of the branch vector (the vector
joining the center of the particle to the point of contact) and the
sum extends over all the contacts $c$ among all particles inside
the representative volume element. The structural anisotropy and mean
coordination number are extracted from the fabric tensor, defined as 
\begin{equation}
  F_{\alpha\beta} =
  \frac{1}{N_c} \sum\limits_{c \in V} n_\alpha^c n_\beta^c\quad,
\end{equation}
were $\vec n$ is the normal vector joining the centers of two
contacting particles, and the sum extends over the contacts $c$. The
structural anisotropy is $a=2(F_2 - F_1)$, where $F_i$ is the $i-$th
eigenvalue of the fabric tensor.

Finally, we define a rotational threshold coefficient $\mu_c$ to
simulate the rotation resistance of complex shapes as follows: If the
total torque $|\vec T_i|$ on the particle $i$ is less than $\mu_c R_i
\sum |\vec F_{n,i}|$, where $\sum |\vec F_{n,i}|$ is the sum of the
normal forces on particle $i$, then the total torque and the angular
velocity are set to zero; otherwise they keep their current values. By
this way it is possible to gradually hinder the rotation of the grains
by changing the parameter $\mu_c$. Since the sum of normal forces
magnitudes depends on the number in contacts, this mechanism could
reflect the interlocking of more general shapes, if rotation
resistance is associated with geometric interlocking.

\section{Results and Discussions}\label{sec:results}
\begin{figure}[!ht]
  \centering
  \includegraphics[scale=0.4]{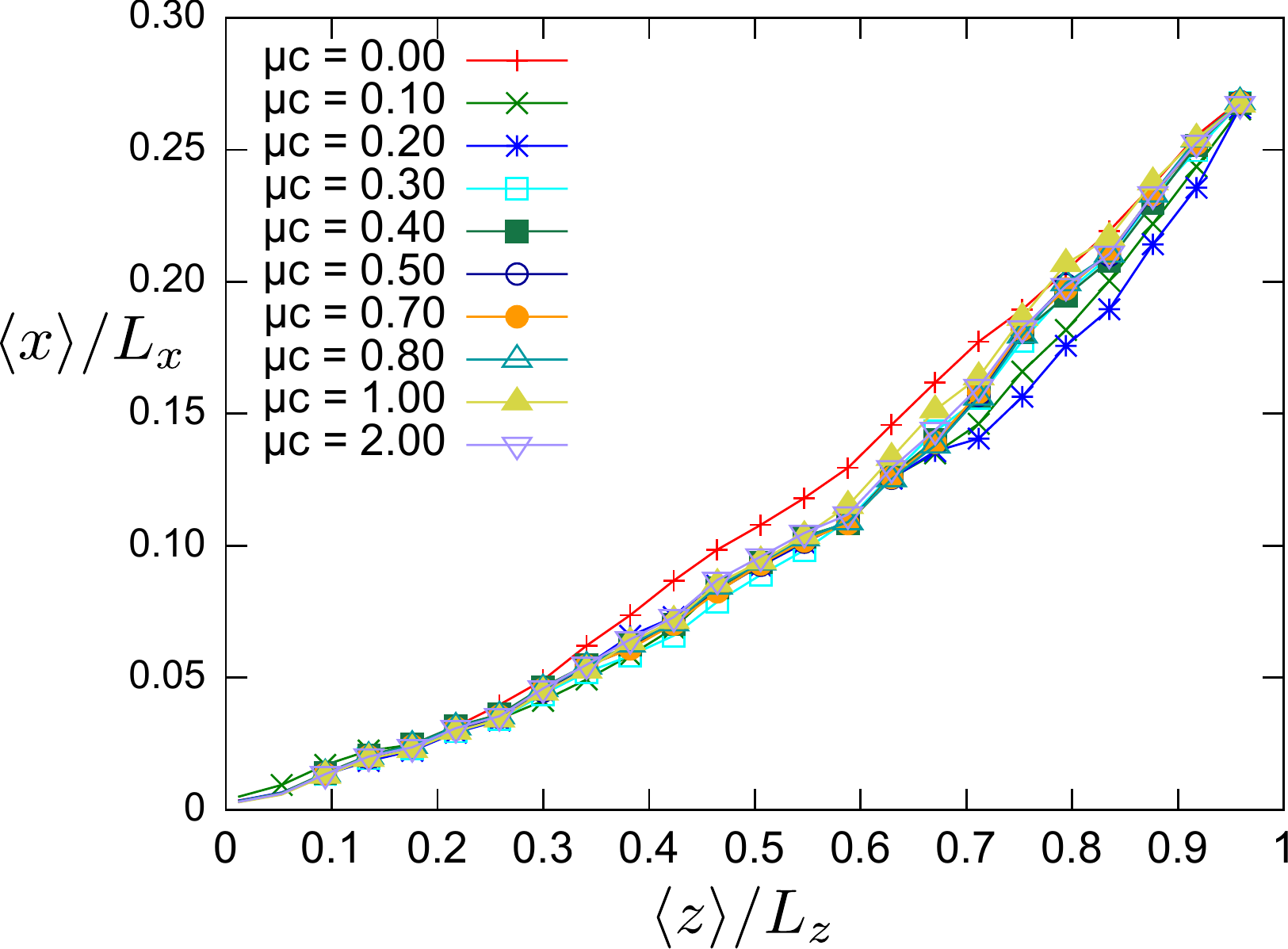}
  \caption{(Color online) Mean normalized horizontal translation as a
    function of the normalized height, for several values of the 
    friction threshold $\mu_c$. $L_x$ and $L_z$ are the horizontal and
    vertical size of the sample, respectively.}\label{fig:meanNetTrans}
\end{figure}

We have performed 2D simulations of periodic simple shear tests on a
poly-disperse sample of 1024 discs. The radius of each particle is
computed as~\cite{poeschelBook}
\begin{equation}
  \label{eq:radiusDistro} R = \frac{R_{\rm min} R_{\rm max}}{R_{\rm
      max} - z(R_{\rm max} - R_{\rm min})},
\end{equation} 
where $z \in [0, 1]$ is a random number.  This particular form is
uniform in \textit{mass} (no size dominates the mass of the system),
and generates more small particles than larger ones, as observed in
natural soils. In contrast, the typical procedure of choosing the
\textit{radius} uniformly as
\begin{equation}
  \label{eq:radiusDistroUniform} R = R_{\rm min} + z(R_{\rm max} -
  R_{\rm min}),
\end{equation} 
will generate almost the same number of small and large particles and
concentrate the mass of system on the larger ones. In the present
simulations, the ratio $R_{\rm max} : R_{\rm min} = 4:1$. The
inter-particle static friction is 0.4. The rolling friction is
0.1. The full sample has been sheared (Figure~\ref{fig:meanNetTrans}),
assuring that no localization band has appeared. In the following,
numerical quantities like $1.23(4)$ should be interpreted as $1.23(4)
= 1.23 \pm 0.04$, i.e. the error is in the last digit, where error
bars correspond to three times the standard deviation. We checked that
$\mu_c \simeq 2$ hinders the rotations almost completely, therefore
there is no need to use larger values for this parameter.

\subsection{Macroscopic internal friction}
\begin{figure}[!ht]
  \centering
  \includegraphics[scale=0.4]{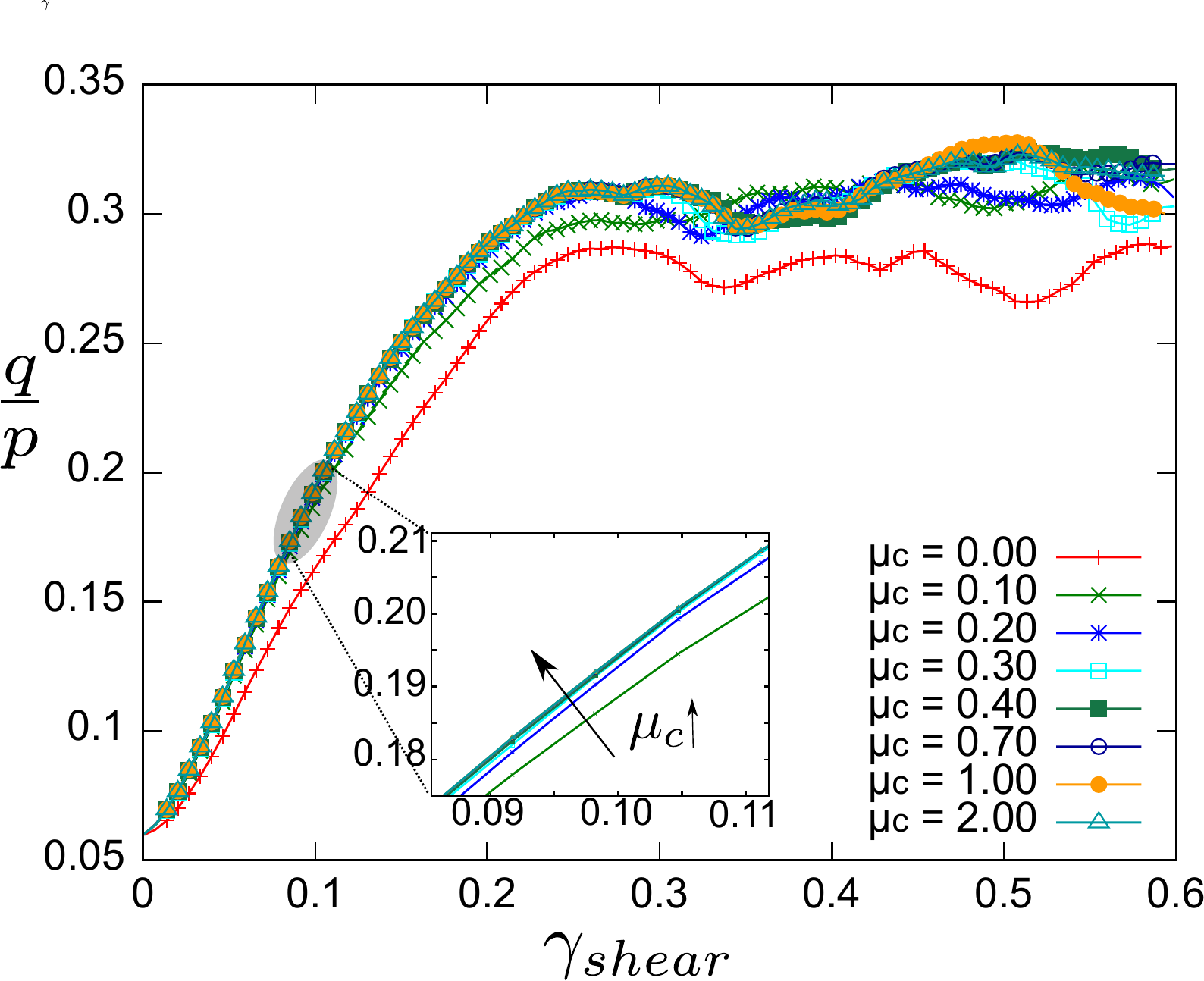}
  \caption{(Color online) $q/p$ as a function of the shear deformation
    $\gamma_{shear} = \Delta x/L_z$, where $\Delta x$ is the
  horizontal translation of the top wall, for several values of the
  threshold. The inset shows a zoom of the approximately linear
  growing regime. }\label{fig:qpVSeq-def}
\end{figure}

The ratio $q/p$ as a function of the shear deformation $\epsilon_q$ is
shown in Figure~\ref{fig:qpVSeq-def}. First, the ratio grows almost
linearly, until it gets the steady state, where it keeps approximately
the same mean value.  The slope at the growing part represents
a transient sample hardness. This hardness increases with the
rotational threshold $\mu_c$.

\begin{figure}[!ht]
  \centering
  \includegraphics[scale=0.4]{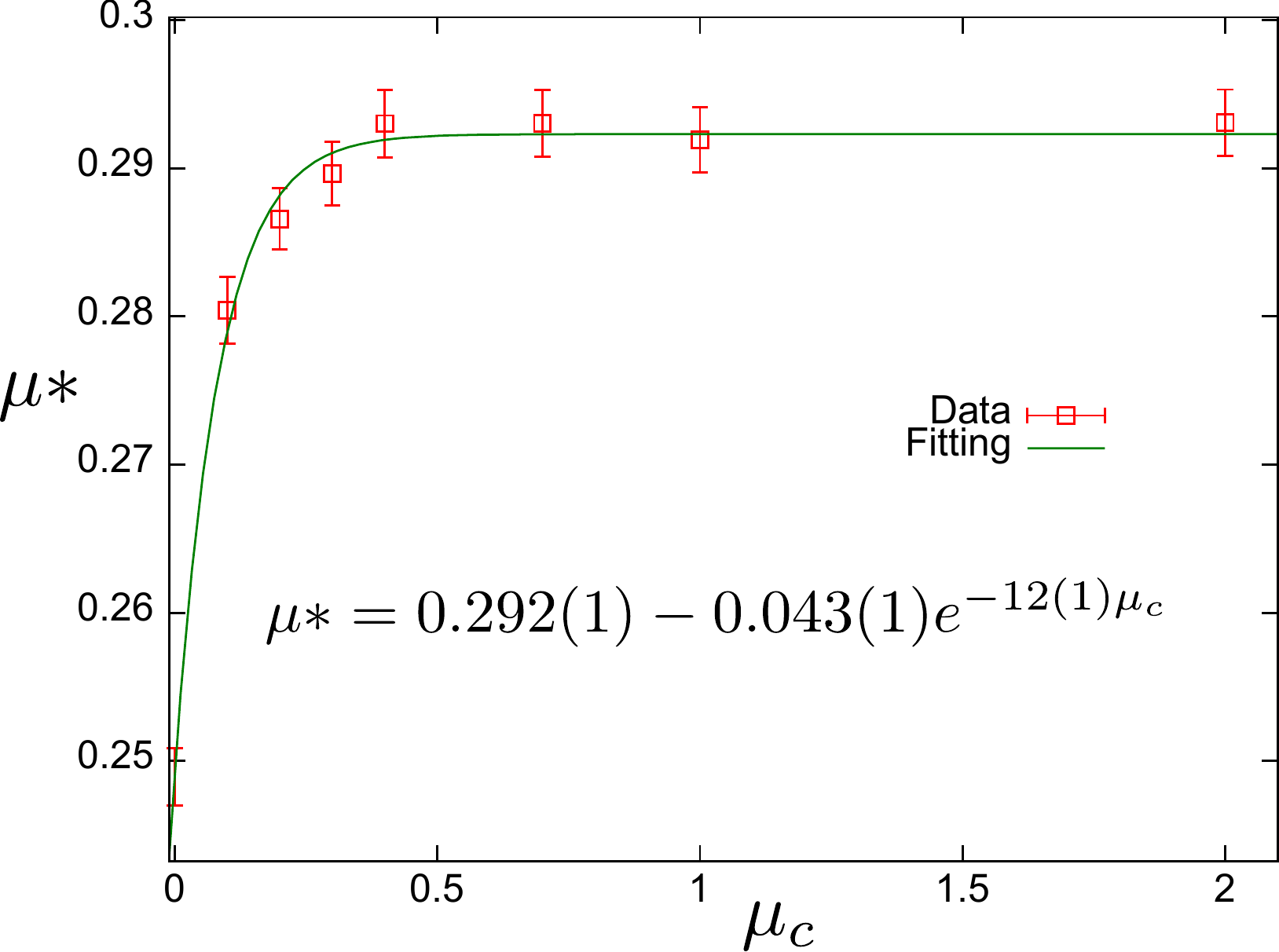}
  \caption{(Color online) Internal friction $\mu*$ for several values of
    the threshold.}\label{fig:internalFriction-def}
\end{figure} 

As we stated before, the threshold $\mu_c$ represents a difficult for
the particles to rotate, analogous to the geometrical interlocking
occurring in more complex-shaped bodies. It is well
known~\cite{azema2007ForceTrans} that irregular bodies has a large
macroscopic internal friction angle $\mu*=\tau/\sigma_{\rm conf}$,
where $\tau$ is the shear stress and $\sigma_{\rm conf}$ is the
confining pressure, because of geometric interlocking. The shear
strength $\mu*$ of the sample increases with increasing $\mu_c$
(Figure~\ref{fig:internalFriction-def}).  This support the
interpretation of $\mu_c$ as quantifying the geometric interlocking of
more complex shapes. The internal macroscopic friction $\mu*$ grows
like a saturated exponential with the form \linebreak $\mu* = 0.292(1)
- 0.043(1)e^{-12(1)\mu_c}$. If $\mu_s$ increases, the final value of
$\mu*$ (equal to $0.292(1)$ in this case) also increases. Therefore,
by tuning of $\mu_s$ and the threshold $\mu_c$, it is possible to
model samples of spheres with very high values of $\mu*$.

\subsection{Distributions of Forces and Torques}
\begin{figure}[!ht]
  \centering
  \includegraphics[scale=0.4]{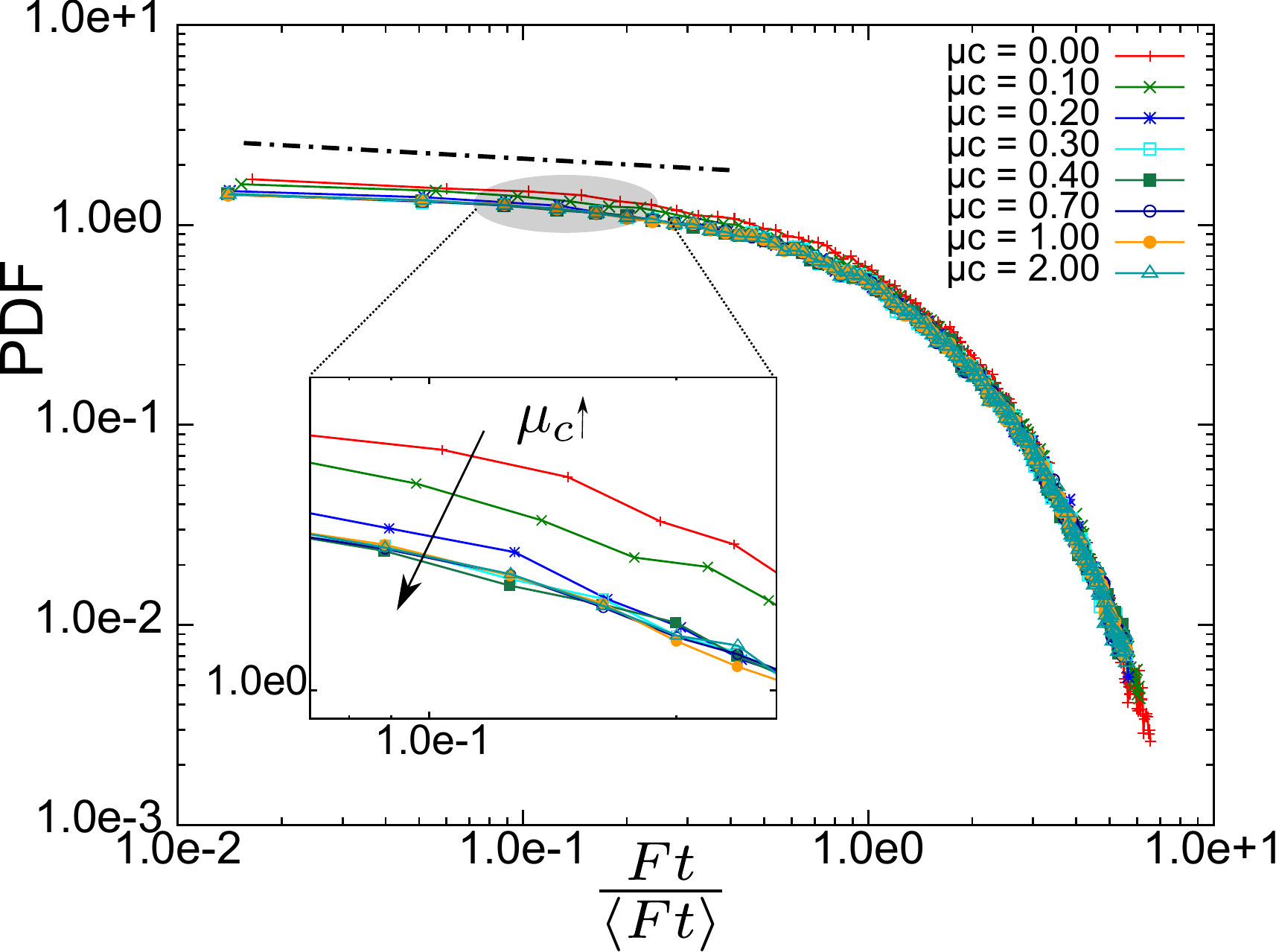}\\
  \includegraphics[scale=0.4]{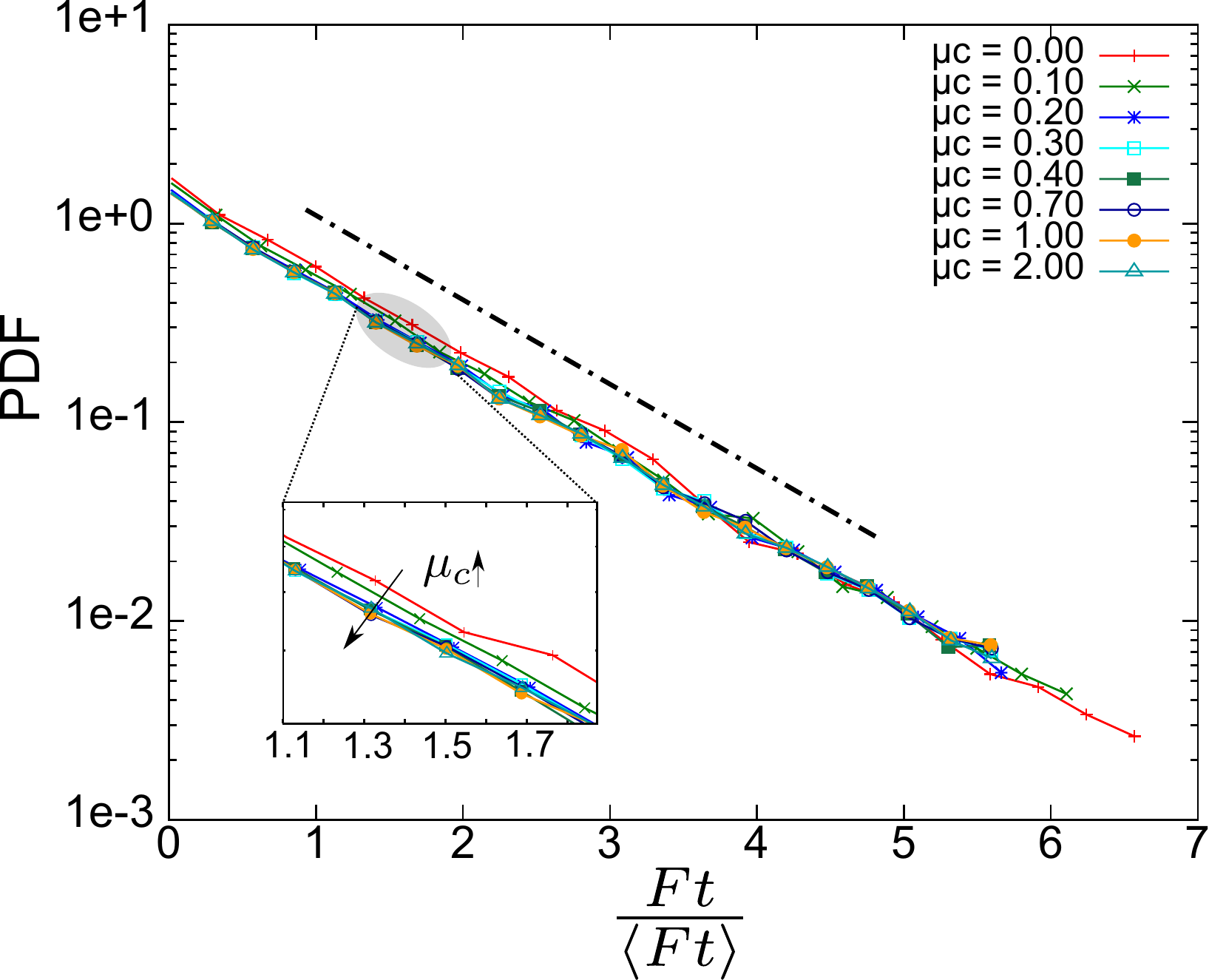}
  \caption{(Color online) Distribution of tangential forces at each contact,
    for several thresholds. Insets: Zoom on given ranges. (Top)
    Log-log, (Bottom) Linear-Log }\label{fig:ftDistro-def}
\end{figure}

\begin{figure}[!ht]
  \centering
  \includegraphics[scale=0.4]{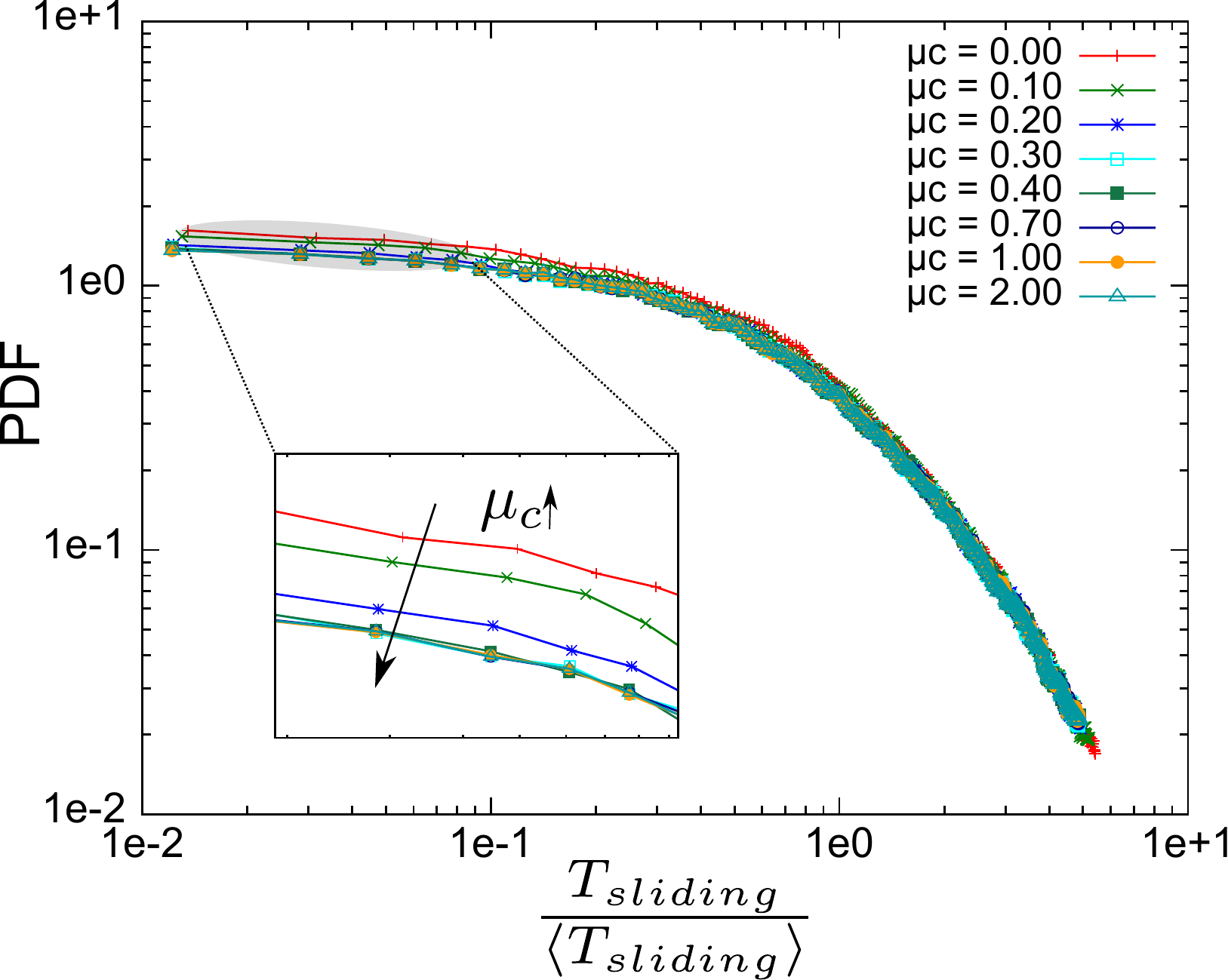}
  \caption{(Color online) Sliding torque distributions for several
    thresholds. Inset: zoom in the range of small
    forces}\label{fig:PTc-def}
\end{figure}

The normalized distributions of forces and torques are presented in
Figures~\ref{fig:ftDistro-def} and \ref{fig:PTc-def} for several
values of the rotational coefficient $\mu_c$. Unexpectedly, it barely
affects the general shape of the distribution of forces and torques in
the sample.

It has been shown that normal forces, tangential forces and even the
torques distribute as a power law for ranges smaller than the mean
and as an exponential for values larger than the
mean~\cite{estradaRolling, radjai1996force}, 
\begin{equation}
 \label{eq:forceDistro}
 P(f) \propto 
 \begin{cases}
   \left(  \frac{f}{\mean{f}}  \right)^\alpha, & f < \mean{f}, \\
   e^{\beta[1 - f/\mean{f}]}, & f > \mean{f}\quad.
 \end{cases}
\end{equation} 
For all the values of $\mu_c$, the exponents $\alpha$ and $\beta$ are
almost the same, $\alpha=-0.16(1)$, while for larger
forces, $\beta=0.93(2)$. Therefore, the forces' and torques'
magnitudes distributions are barely affected by the threshold,
although the relative importance of each particular distribution is
slightly different.    

\subsection{Mean Coordination number and structural anisotropy}
\begin{figure}[!ht]
  \centering
  \includegraphics[scale=0.4]{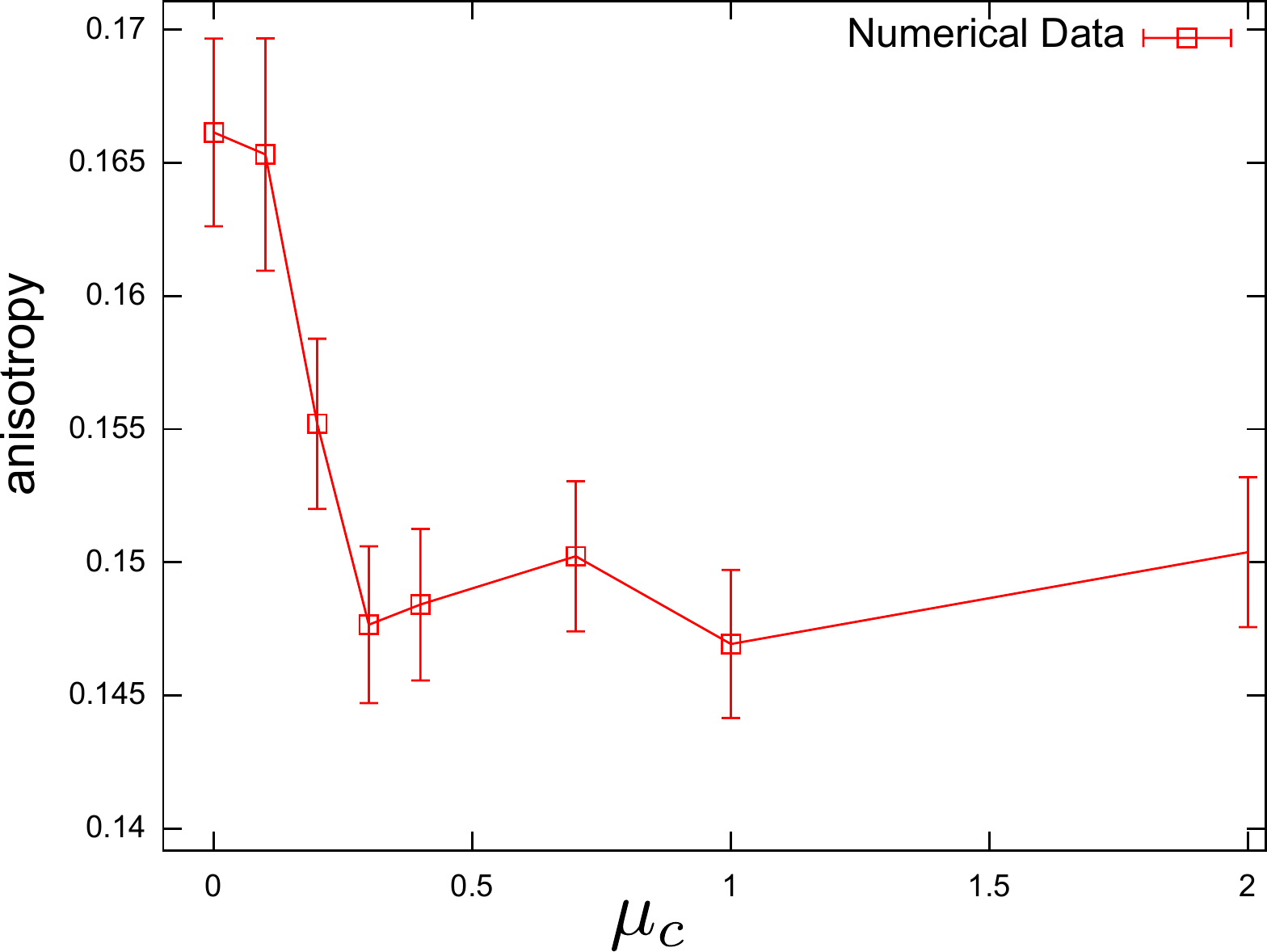}
 \caption{(Color online) Structural anisotropy as a function of the  
    thresholds.}\label{fig:anisotropy-def}.
\end{figure}

The mean coordination number, on one hand, is almost constant for all
the values of $\mu_c$, with a value of $\langle z
\rangle$$=$$4.1(2)$. On the other hand, the structural anisotropy
starts with a small value and 
decays rapidly to almost a constant value for large thresholds
(Fig.~\ref{fig:anisotropy-def}). 

Although the mean contact number is practically the same for all
thresholds, we found that the larger the threshold the smaller the
structural anisotropy. This result agrees
with~\cite{azema2007ForceTrans}, where discs' and pentagons' samples
were compared in biaxial tests and it was found that discs' samples
have higher structural signed anisotropies and smaller strengths,
because of smaller force anisotropy. If the threshold $\mu_c$ can be
associated with shape, then larger values of $\mu_c$ should reflect
more irregular shape, higher shear strength and smaller structural
anisotropy, as found. The higher shear strength with smaller
structural anisotropy can be explained by means of the forces'
anisotropies, that reflects how the forces magnitudes are distributed
along the contact network. Future works could check that increasing
$\mu_c$ enhances the force anisotropies and thus the shear
strength. Since the structural anisotropy decreases rapidly to a
constant value, the parameter $\mu_c$ is shown to captures some but
not all the effects of irregular shapes. For increasing values of
$\mu_c$, the connectivity of the contact network
is kept constant in average, while the quantities that are carried
over this network are changing,  since $\langle z \rangle$ is
almost constant while the structural anisotropy decreases for
increasing $\mu_c$.    

\section{Conclusions}\label{sec:conclu}
In this work we studied the effect of a gradual hindering of the
rotations on the critical state of a two-dimensional dry soil of
poly-disperse spheres with size span 4:1. For this purpose, we have
simulated a simple shear test for several values of a rotational
threshold friction coefficient $\mu_c$, acting as follows: the
rotation of a sphere is allowed only if the magnitude of the total
torque on the sphere is larger than the sum of the magnitudes of the
normal forces at all contacts times the threshold $\mu_c$. So, if
$\mu_c=0$ all torques (and its respective effect on rotations) are
allowed, while if $\mu_c \to \infty$, no rotation occurs. Actually, we
found that values of $\mu_c \simeq 2 $ are enough to hinder almost
completely all rotations.

In order to characterize the effect of $\mu_c$ on the macroscopic
response of the system, five quantities were tracked: the stress ratio
$q/p$, the total internal friction $\mu*$, the distribution of forces
and torques, the mean correlation number $\langle z \rangle$, and the
structural anisotropy of the fabric tensor. We found that the final
stress ratio $q/p$ is slightly increased for larger values of $\mu_c$;
a consistent increment of the sample hardness before the critical
state is reached can be easily observed. The normalized distribution
of forces' and torques' magnitudes are barely affected in terms of its
characteristic exponents, although is relative importance seems to
decrease for higher values of $\mu_c$. The mean correlation number is
barely affected by the rotational threshold $\mu_c$, which means that
the connectivity of the contact network is not changed.  In contrast,
the macroscopic friction coefficient $\mu*$ does increase with
$\mu_c$, like a saturated exponential, while the structural anisotropy
decreases. This last result agrees with previous hindering strategies
at the level of contact (instead of at the level of particles as
here)~\cite{azema2007ForceTrans}. This results enhances the
interpretation of $\mu_c$ as characterizing the effect of complex
shapes and its natural geometric interlocking. Further works could
check the evolution of the force anisotropies in order to explain the
increasing shear strength with decreasing structural anisotropies. We
expect the force anisotropies to increase with $\mu_c$. The threshold
$\mu_c$ captures some but not all the features of more complex shapes.

Different values of the static friction coefficient between grains
produce different values for the global internal friction coefficient
$\mu*$, and the value of the later can be further increased by means
of the threshold $\mu_c$. This relationship can be exploited to
simulate materials composed of spheres and with very high strengths,
but further calibration with models including irregular shapes is
needed.

This work is a contribution to a better understanding of the critical
state and the role of the rotational degrees of freedom. The exact
dependence for different macroscopic parameters, for different
macroscopic conditions, the role of forces anisotropies, and its
quantitative comparison with more complex shape materials are topics
of future work.

\section{Acknowledgments}
We thank the CEiBA complex systems research center for financial support,
and Nicol\'as Estrada and Mauricio Boton for helpful discussions.





\bibliographystyle{elsarticle-num}
\bibliography{biblioGeneral}







\end{document}